\documentclass[12pt]{article}
\usepackage{euscript,amsmath, amssymb, amsfonts, color}

\newcommand{\half}{\frac{1}{2}}

\newcommand{\R}{{R}}
\newcommand{\be}{\begin{equation}}
\newcommand{\ee}{\end{equation}}
\newcommand{\bee}{\begin{eqnarray}}
\newcommand{\eee}{\end{eqnarray}}
\newcommand\nn{\nonumber \\}

\newcommand{\di}{\frac {\partial} {\partial x_i}}
\newcommand\defeq{\stackrel {def}{=}}
\newcommand{\x}{{\vec x}}
\newcommand{\yy}{{\vec y}}

\newcommand{\vv}{{\vec v}}

\newcounter{theorem}
\newcommand{\theorem}{\par\refstepcounter{theorem}
           {\bf Theorem 
           \arabic{theorem}. }}

\newcounter{lemma}

\newcounter{proposition}

\newcounter{definition}
\newcommand{\definition}{\par\refstepcounter{definition}
           {\bf Definition 
           \arabic{definition}. }}

\font\frtnfr=eufm10   scaled\magstep1
\font\twlfr=eufm10
\font\tenfr=eufm10

\newfam\frfam
\textfont\frfam=\frtnfr
\scriptfont\frfam=\twlfr
\scriptscriptfont\frfam=\tenfr
\def\fr{\fam\frfam}

\font\frtnopen=msbm10  scaled\magstep2
\font\twlopen=msbm10
\font\tenopen=msbm10

\newfam\openfam
\textfont\openfam=\frtnopen
\scriptfont\openfam=\twlopen
\scriptscriptfont\openfam=\tenopen
\def\open{\fam\openfam}

\font\frtnsf = cmss12 scaled\magstep1
\font\twlsf = cmss10
\font\tensf = cmss9

\newfam\Scfam
\textfont\Scfam = \frtnsf
\scriptfont\Scfam = \twlsf
\scriptscriptfont\Scfam = \tensf

\begin{document}
\renewcommand{\theequation}{\arabic{equation}}
\bibliographystyle{nphys}


\sloppy
\title
 {
      The Number of Supertraces on the Superalgebra of Observables of
      Rational Calogero Model based on the Root System
 }
\author
 {
 S.E.~Konstein\thanks{E-mail: konstein@lpi.ru}
 \thanks{
               This work was supported
               by the RFBR (grant No.~08-02-01118 (S.K.)),
               and by the grant LSS-1615.2008.2 (S.K.).
 }
\\ {\small
               \phantom{uuu}} \\ {\small I.E.Tamm Department of
               Theoretical Physics,} \\
               {\small P. N. Lebedev Physical
               Institute,} \\ {\small 119991, Leninsky Prospect 53,
               Moscow, Russia.}
\\
\phantom{uuu} \ \\
and
\\
\phantom{uuu} \ \\
 R.~Stekolshchik\thanks{E-mail: rs2@biu.013.net.il}
\\ {\small
               \phantom{uuu}} \\ {\small ECI Telecomm  Ltd.,} \\
               {\small Osivim 30, Petah-Tikva, Israel. }}

\date{}
\maketitle
\begin{abstract}
In the Coxeter group $W(\R)$ generated by the root system $\R$, let
$Q(\R)$ be the number of conjugacy classes having no eigenvalue $-1$.
The superalgebra $H_{W(\R)}$ of observables of the rational Calogero
model based on the root system $\R$ possesses $Q(\R)$ supertraces.
The numbers $Q(\R)$ are determined for all irreducible root systems
(hence for all root systems).
\end{abstract}

\newpage

\section{The superalgebra of observables}
The superalgebra $H_{W(\R)(\nu)}$ of observables
of the rational Calogero model based on the root system $\R$
is defined in the following way.

For any nonzero $\vv \in V={\open R}^N$ define the reflections
$R_\vv$ as follows:
\be\label{ref}
R_\vv (\x)=\x -2 \frac {(\x,\,\vv)} {(\vv,\,\vv)} \vv \qquad
\mbox{ for any }\x \in V.
\ee
Here $(\cdot,\cdot)$ stands for the inner product in
$V$, i.e., $(\x,\,\yy)=\sum_{i=1}^N x_i y_i$, where the $x_i$ are
the coordinates of vector $\x$: $x_i\defeq (\x,\,\vec e_i)$,
and the vectors $\vec e_i$ constitute an orthonormal basis in $V$:
$(\vec e_i,\, \vec e_j)=\delta_{ij}$.
The reflections (\ref{ref}) have the following properties
\be
R_\vv (\vv)=-\vv,\qquad R_\vv^2 =1,\qquad
({R}_\vv (\x),\,\vec u)=(\x,\,{R}_\vv (\vec u)),\quad
\mbox{ for any }\vv,\,\x,\,\vec u\in V.\nonumber
\ee

\definition
{\it The finite set of vectors $\R\subset V$ is a {\it root system}
if $\R$ is ${R}_\vv$-invariant for any $\vv \in \R$ and
the group $W(\R)$ generated by all reflections ${R}_\vv$ with
$\vv \in \R$ (called a Coxeter group) is finite.}

In this paper, we consider noncrystallographic irreducible root systems
($H_3$, $H_4$ and $I_2(n)$) together with crystallographic ones
($A_n$, $B_n$, $C_n$, $D_n$, $E_n$, $F_4$, $G_2$) \cite{NB,JH}.

Let ${\cal H}^\alpha$ ($\alpha=0,1$) be two copies of $V$ with
orthonormal bases $a^\alpha_i$ ($i=1,\,...\,,\,N$), respectively.
For every vector $\vv=\sum_{i=1}^N v_i{\vec e_i}\in V$ let
$v^\alpha\in {\cal H}^\alpha$
be the vectors $v^\alpha=\sum_{i=1}^N v_i a^\alpha_i$,
so the bilinear forms on 
${\cal H}^0\oplus {\cal H}^1$
can be defined as
$
(x^\alpha,\,y^\beta)=(\vec x,\, \vec y),
$
where $\vec x,\,\vec y \in V$ and $x^\alpha,\,y^\alpha \in {\cal
H}^\alpha$ are their copies.
The reflections $R_\vv$ act on ${\cal H}^\alpha$ as follows
\be
R_\vv(h^\alpha)=h^\alpha -
2\frac{(h^\alpha,\,v^\alpha)}{(\vv,\,\vv)}v^\alpha,
\qquad \mbox{ for any } h^\alpha\in {\cal H}^\alpha.
\nonumber
\ee
Thus, the $W(\R)$-action on the spaces ${\cal H}^\alpha$
is defined.

Let $\nu$ be a set of constants $\nu_\vv$ with $\vv\in\R$ such that
$\nu_\vv=\nu_{\vec w}$ if $R_\vv$ and $R_{\vec w}$ belong to
one conjugacy class of $W(\R)$.

\definition
{\it
$H_{W({R})}(\nu)$ is the associative algebra
of polynomials in the $a^\alpha_i$ with coefficients
in the group algebra ${\open C}[W(\R)]$ subject to the relations
\bee
gh^\alpha=g(h^\alpha)g \quad \mbox{ for any } g\in W(\R),
                  \quad \mbox{ and } h^\alpha \in {\cal H}^\alpha \nn
\label{rel}
  [ h_1^\alpha, h_2^\beta] = \varepsilon^{\alpha\beta}
       \left((\vec h_1,\, \vec h_2)+
       \sum_{\vv\in\R} \nu_\vv
\frac {(\vec h_1,\,\vv)(\vec h_2,\,\vv)}{(\vv,\,\vv)}R_\vv\right)
\mbox{ for any  $h_1^\alpha$, $h_2^\alpha\, \in {\cal H}^\alpha$},
\eee
where $\varepsilon^{\alpha\beta}$
is the antisymmetric tensor, $\varepsilon^{01}=1$.
}

This algebra 
has faithful representation
via Dunkl differential-difference operators \cite{Dunkl} acting on
the space of smooth functions on $V$. Namely, let
\be
D_i=
\di +\half \sum_{\vv\in\R} \nu_\vv\frac {v_i} {(\x,\,\vv)} (1-R_\vv)
\nonumber
\ee
and
\cite{Poly, BHV}
\be
\label{aa}
a_i^\alpha =\frac 1 {\sqrt{2}} (x_i + (-1)^\alpha D_i),\quad
\alpha =0,1.
\ee

The reflections $R_\vv$ transform the deformed creation and annihilation
operators
(\ref{aa})
as vectors:
\be
R_\vv a_i^\alpha = \sum_{j=1}^N \left(\delta_{ij} - 2
\frac {v_i v_j}{(\vv,\,\vv)}\right)a_j^\alpha  R_\vv.
\nonumber
\ee
Since $[D_i,\, D_j]=0$ \cite{Dunkl}, it follows that
\be
[a^\alpha_i, a^\beta_j] = \varepsilon^{\alpha\beta}
\left(\delta_{ij}+
\sum_{\vv\in\R} \nu_\vv \frac {v_i v_j}{(\vv,\,\vv)}R_\vv\right),
\nonumber
\ee
which manifestly coincides with (\ref{rel}).

The commutation relations (\ref{rel}) suggest
to define the {\it parity} $\pi$ by setting:
\be
\pi (a_i^\alpha)=1\ \mbox{ for any }\alpha,\,i,
\qquad \pi(g)=0 \ \mbox{ for any } g\in W(\R)
\nonumber
\ee
and consider $H_{W({R})}(\nu)$ as a superalgebra.

We say that $H_{W({R})}(\nu)$ is {\it the superalgebra of
observables of the Calogero model based on the root system $\R$}.

Obviously, ${\open C}[W(\R)]$
is a subalgebra of $H_{W({R})}(\nu)$.

Observe an important property of superalgebra
$H_{W({R})}(\nu)$: the Lie superalgebra of its inner
derivations\footnote{Let ${\cal A}$ be arbitrary associative
superalgebra. Then, the operators ${\cal D}_x$
which act on ${\cal A}$ via
${\cal D}_x(y)=[x,\,y\}$ (supercommutator) constitute
the Lie superalgebra of {\it inner}
derivations.} contains  ${\fr sl}_2$ generated by
\be
T^{\alpha\beta}=\half \sum_{i=1}^N
\left\{a_i^\alpha,\,a_i^\beta\right\}
\nonumber
\ee
which commute with ${\open C}[W(\R)]$, i.e.,
$[T^{\alpha\beta},\,R_\vv]=0$,
and act on $a_i^\alpha$ as on ${\fr sl}_2$-vectors:
\be\label{sl2vec}
\left[T^{\alpha\beta},\,a_i^\gamma\right]=
\varepsilon^{\alpha\gamma}a_i^\beta +
\varepsilon^{\beta\gamma}a_i^\alpha.
\ee

The restriction of operator
$T^{01}$ in the representation (\ref{aa}) on the subspace
of $W(\R)$-invariant functions on $V$
is a second-order differential operator which is the well-known
Hamiltonian of the rational Calogero model \cite{Cal} based on the
root system $\R$ \cite{OP}.
One of the relations (\ref{sl2vec}), namely,
$[T^{01},\,a_i^\alpha]=
-(-1)^\alpha a_i^\alpha$, allows one to find the wave functions
of
the equation $T^{01}\psi =\epsilon \psi$
via the usual Fock procedure with
the vacuum $ |0\rangle$ such that
$a_i^0|0\rangle$=0 $\mbox{ for any } i$ \cite{BHV}.
After $W(\R)$-symmetrization these wave functions
become the wave functions of Calogero Hamiltonian.


\section{Supertraces on $H_{W({R})}(\nu)$}\label{trace}

\definition
{\it Any linear complex-valued function $str(\cdot)$ on the superalgebra
${\cal A}$ such that
\bee
str(fg)&=&(-1)^{\pi(f)\pi(g)}str(gf)
\nonumber
\eee
for any $f,g \in
{\cal A}$ with definite parity $\pi(f)$ and $\pi(g)$ is called a
supertrace.}

The restriction of every supertrace on ${\open C}[W(\R)]$
is completely determined by its values on
$W(\R)\subset {\open C}[W(\R)]$
and the function $str$ is a central function on $W(\R)$, i.e.,
it is constant on the conjugacy classes.

Let
us introduce the grading $E$ on the vector space of ${\open C}[W(\R)]$.
Consider the subspace ${\cal E} (g)\subset V$:
\be
{\cal E} (g) =\{x\in V\mid\quad gx=-xg \}\mbox{ for }
g\in W(\R).
\nonumber
\ee
Set
\be
E(g)=dim\,{\cal E}(g).
\nonumber
\ee

The following theorem was proved in \cite{-1}

\theorem\label{theor}
{\it The number of linearly
independent supertraces $Q(\R)$ on $H_{W({R})}(\nu)$
does not depend on $\nu$ and
is equal to
the number of conjugacy classes in $W(\R)$ with $E(g)=0$.}

This Theorem helps to find the number $Q(\R)$ for an arbitrary
root system $\R$.

Theorem \ref{theor} implies, evidently, the following statement

\theorem
\be
Q(\R_1 \oplus \R_2) = Q(\R_1) Q(\R_2).
\nonumber
\ee

Therefore, the problem of finding $Q(\R)$ is reduced to the problem
of finding $Q(\R)$ for irreducible root systems $\R$.

\section{The numbers $Q(\R)$ for irreducible root systems}

\bigskip
{
\centerline
{\renewcommand{\arraystretch}{0}%
\begin{tabular}{|c|c|c|}
  \hline
\phantom {\small uu} &\phantom {\small uu}&\phantom {\small uu}\\
  $\R$ & $Q(\R)$ & for the proof, see  \\
\phantom {\tiny uu} &\phantom {\tiny uu}&\phantom {\tiny uu}\\
  \hline
\rule{0pt}{2pt}&&\\
  \hline
\phantom {\small uu} &\phantom {\small uu}&\phantom {\small uu}\\
  $A_{n-1}$ &
         \begin{tabular}{l}
  the number of partitions of $n$\\
  into a sum of odd positive integers
         \end{tabular}
  & \cite{KV} \\
\phantom {\tiny uu} &\phantom {\tiny uu}&\phantom {\tiny uu}\\
   \hline
\phantom {\small uu} &\phantom {\small uu}&\phantom {\small uu}\\
  $B_{n},C_{n},BC_{n}$ &
         \begin{tabular}{l}
  the number of partitions of $n$\\
  into a sum of positive integers
         \end{tabular}
  & \cite{root} \\
\phantom {\tiny uu} &\phantom {\tiny uu}&\phantom {\tiny uu}\\
   \hline
\phantom {\small uu} &\phantom {\small uu}&\phantom {\small uu}\\
  $D_{n}$ &
           \begin{tabular}{l}
   the number of partitions of $n$\\
   into a sum of positive integers\\
   with even number of  even integers
          \end{tabular}
  & \cite{root} \\
\phantom {\tiny uu} &\phantom {\tiny uu}&\phantom {\tiny uu}\\
   \hline
\phantom {\small uu} &\phantom {\small uu}&\phantom {\small uu}\\
  $E_6$ & 9 & Appendix \ref{6} \\
  \phantom {\tiny uu} &\phantom {\tiny uu}&\phantom {\tiny uu}\\
   \hline
\phantom {\small uu} &\phantom {\small uu}&\phantom {\small uu}\\
  $E_7$ & 12 & Appendix \ref{7} \\
  \phantom {\tiny uu} &\phantom {\tiny uu}&\phantom {\tiny uu}\\
   \hline
\phantom {\small uu} &\phantom {\small uu}&\phantom {\small uu}\\
$E_8$ & 30 & Appendix \ref{8} \\
\phantom {\tiny uu} &\phantom {\tiny uu}&\phantom {\tiny uu}\\
   \hline
\phantom {\small uu} &\phantom {\small uu}&\phantom {\small uu}\\
  $F_4$ & 9 & Appendix \ref{4} \\
\phantom {\tiny uu} &\phantom {\tiny uu}&\phantom {\tiny uu}\\
   \hline
\phantom {\small uu} &\phantom {\small uu}&\phantom {\small uu}\\
  $G_2$ & 3 & \cite{i2} \\
\phantom {\tiny uu} &\phantom {\tiny uu}&\phantom {\tiny uu}\\
   \hline
\phantom {\small uu} &\phantom {\small uu}&\phantom {\small uu}\\
  $H_3$ & 4 & Appendix \ref{3} \\
\phantom {\tiny uu} &\phantom {\tiny uu}&\phantom {\tiny uu}\\
   \hline
\phantom {\small uu} &\phantom {\small uu}&\phantom {\small uu}\\
  $H_4$ & 20 & Appendix \ref{H4} \\
\phantom {\tiny uu} &\phantom {\tiny uu}&\phantom {\tiny uu}\\
   \hline
\phantom {\small uu} &\phantom {\small uu}&\phantom {\small uu}\\
  $I_2(n)$ & {\Large $\left[\frac {n+1} 2 \right]$} & \cite{i2} \\
\phantom {\ uu} &\phantom {\small uu}&\phantom {\tiny uu}\\
   \hline
\end{tabular}%
}
}


\bigskip
\newpage
\section*{Appendices}
\bigskip



\setcounter{equation}{0} \def\theequation{A\arabic{appen}.\arabic{equation}}

\newcounter{appen}
\newcommand{\appen}[1]{\par\refstepcounter{appen}
{\par\bigskip\noindent\large\bf Appendix \arabic{appen}. \medskip }{\bf \large{#1}}}

\renewcommand{\subsection}[1]{\refstepcounter{subsection}
{\bf A\arabic{appen}.\arabic{subsection}. }{\ \bf #1}}
\renewcommand\thesubsection{A\theappen.\arabic{subsection}}
\makeatletter \@addtoreset{subsection}{appen}

\renewcommand{\subsubsection}{\par\refstepcounter{subsubsection}
{\bf A\arabic{appen}.\arabic{subsection}.\arabic{subsubsection}. }}
\renewcommand\thesubsubsection{A\theappen.\arabic{subsection}.\arabic{subsubsection}}
\makeatletter \@addtoreset{subsubsection}{subsection}


\appen{$E_6$}\label{6}

The conjugacy classes of the Weyl group $E_6$
are described in Table 9 of \cite{carter}.

The following 9 classes have not the root $-1$:
$$
   \phi,
\    A_2,
\   A_2^2,
  \   A_4,
\   D_4(a_1),
\   A_2^3,
\   E_6,
\   E_6(a_1),
\   E_6(a_2).
$$

\appen{$E_7$}\label{7}

The conjugacy classes of the Weyl group $E_7$
are described in Table 10 of \cite{carter}.

The following 12 classes have not the root $-1$:
$$
   \phi,
\    A_2,
\   A_2^2,
  \   A_4,
\   D_4(a_1),
\   A_2^3,
\  A_4 \times A_2,
\  A_6,
\   D_6(a_1),
\   E_6,
\   E_6(a_1),
\   E_6(a_2).
$$

\appen{$E_8$}\label{8}

The conjugacy classes of the Weyl group $E_8$ are described in Table 11 of \cite{carter}.

The following 30 classes have not the root $-1$:
\begin{eqnarray*}
&&
\phi,
\    A_2,
\   A_2^2,
\   A_4,
\   D_4(a_1),
\   A_2^3,
\  A_4 \times A_2,
\  A_6,
\  D_4(a_1) \times A_2,
\   D_6(a_1),
\\
&&
E_6,
\   E_6(a_1),
\   E_6(a_2),
\   A_2^4,
\   A_4^2,
\   A_8,
\   D_4(a_1)^2,
\   D_8(a_1),
\   D_8(a_3),
\\
&&
E_6 \times A_2,
\   E_6(a_2) \times A_2,
\   E_8,\  E_8(a_i)\  (i = 1,\,...\,8).
\end{eqnarray*}

\appen{$F_4$}\label{4}

The conjugacy classes of the Weyl group $F_4$
are described in Table 8 of \cite{carter}.

The following 9 classes have not the root $-1$:
$$\phi,\ A_2, \ \tilde A_2,\ B_2,\ A_2 \times \tilde A_2,\ D_4(a_1),
\ B_4,\ F_4,\ F_4(a_1).$$

\appen{$H_3$}\label{3}

Let $k=(\sqrt {5} +1)/2$.
Then the reflections
\begin{eqnarray*}
a
& =&
\left(
\begin{array}{ccc}
1& 0& 0 \\
0 &-1& 0 \\
0 &0& 1
\end{array}
\right)
\\
b& =&\frac 1 2
\left(
\begin{array}{ccc}
1& k& k-1 \\
k &1-k& -1 \\
k-1 &-1& k
\end{array}
\right)
\\
c & =&
\left(
\begin{array}{ccc}
-1& 0& 0 \\
0 &1& 0 \\
0 &0& 1
\end{array}
\right)
\\
\end{eqnarray*}
corresponding to the roots $\vec e_2$,
$\frac 1 2 (-\vec e_1 + k\vec e_2 + k^{-1}\vec e_3)$
and $\vec e_1$
satisfy the relations
$$a^2 = b^2 = c^2 = 1, \ \ (ab)^5 = (bc)^3 = (ac)^2 = 1$$
and define the Coxeter group $H_3$.

As
$H_3 = S^e_5 \times C_2$ (See \cite{NB},  Ch.VI, Sect. 4, Ex.11 d),  p. 284;
\cite{h4}, p.160 and references there),
where $S^e_5$ is the group of even permutations of 5 elements,
$C_2=\{1,-1\}$,
($|S^e_5| = 60$,  $|H_3| = 120$),
$H_3$ has 10 conjugacy classes, 5 with positive determinant
and 5 with negative one.

The conjugacy classes with positive determinant are
described by their representatives

\bigskip
\centerline{
\begin{tabular}{|l|l|}
   \hline
\phantom {\tiny uu} &\phantom {\tiny uu}\\
The representative & The characteristic polynomial \\
\phantom {\tiny uu} &\phantom {\tiny uu}\\
   \hline
\phantom {\tiny uu} &\phantom {\tiny uu}\\
$ 1$ &  $( 1 - t )^3$
\\
$ ac$  & $( 1 - t )( 1 + t )^2$
\\
$ bc$  & $( 1 - t )( t^2 + t + 1 )$
\\
 $ab$  & $( 1 - t ) [ t^2 + ( 1 - k ) t + 1 ]$
\\
$ abab$  & $( 1 - t ) ( t^2 + k t + 1 )$
\\
\phantom {\tiny uu} &\phantom {\tiny uu}\\
\hline
\end{tabular}
}
\bigskip

Each of this characteristic polynomials has the root $+1$,
and so each characteristic polynomials of any conjugacy class with negative
determinant has the root $-1$.
The conjugacy class with representative $ac$ also has the root $-1$.

So, the number of conjugacy classes without root $-1$ is equal to 4.

\appen{$H_4$}\label{H4}

According to \cite{h4}, all 34 conjugacy classes of $H_4$ are described by
their representatives acting on the space of quaternions
\be\label{lr}
g_{lr}:\ \ x \mapsto l x r^*
\ee
and
\be\label{p}
g^*_{p}:\ \ x \mapsto p x^*.
\ee
and all 25 pairs of unit quaternions $l$ and $r$ and 9 unit quaternions $p$
are listed in Table 3 of \cite{h4}.

Each operator (\ref{p}) has the root $-1$.
Indeed, the equation
$
px^*=-x
$
has nonzero solution
$
x= {-1+p}
$
if $p\ne 1$,
and $x$ is an arbitrary imaginary quaternion if $p=1$.

Each operator (\ref{lr}) has not the root $-1$ if and only if
\be\label{ne}
l_0+r_0\ne 0.
\ee
Indeed, the determinant of the map
$x\mapsto lx+xr$ is equal to $4(l_0 + r_0)^2$.

There are 20 pairs of $l,r$ in Table 3 of \cite{h4}
satisfying the condition (\ref{ne}), namely,
$K_i$ with $i$ = 1, 4, 6, 8, 10 -- 25.

\newpage

\end{document}